# Focusing and Energy Dispersion Properties of a Cylindrically Bent Asymmetric Laue Crystal

Peng Qi[a], Xianbo Shi[b], Nazanin Samadi[c,d], Dean Chapman*[d,e]

[a] Division of Biomedical Engineering, University of Saskatchewan, Saskatoon, Canada
[b] Advanced Photon Source, Argonne National Laboratory, Lemont, Illinois, USA
[c] Department of Physics and Engineering Physics, University of Saskatchewan, Saskatoon, Canada
[d] Canadian Light Source, Saskatoon, Canada
[e] Anatomy, Physiology & Pharmacology, University of Saskatchewan, Saskatoon, Canada

## ABSTRACT

Elastically bent single crystal Laue case diffraction crystals provide interesting new opportunities for imaging and spectroscopy applications. The diffraction properties are well understood, however, the ability to easily model the diffracted beams hinders assessment of the focal, phase and energy dispersive properties needed for many applications. This work begins to collect the elements needed to ray trace diffracted beams within bent Laue crystals for the purpose of incorporation into other powerful ray tracing applications such as SHADOW. Specifically, we address the condition in a bent Laue crystal where a cylindrically bent Laue crystal will focus all the polychromatic diffracted beams at a single location when a specific asymmetry angle condition is met for a target x-ray energy – the so-called 'magic condition'. The focal size of the beam can be minimized, but this condition also results in excellent energy dispersive properties. The conceptual and mathematical aspects of this interesting focusing and energy dispersive phenomenon is discussed.

## 1. INTRODUCTION

X-ray Laue or transmission type monochromators have been used for many types of applications. Their use has typically been limited due to reduced efficiency compared to Bragg or reflection type monochromators. However, in high x-ray energy applications, the small Bragg angle of the incident x-rays to the crystal surface in the Bragg geometry severely limit the acceptance of the monochromator in the diffraction plane. Also, the Laue geometry is better at handling high incident powers that are generated by synchrotron or other high-power x-ray sources.

An example of the Laue and Bragg geometries is shown in Fig. 1. Note in the figures that the lattice planes can be inclined relative to the crystal surfaces and the angle forms what is called the asymmetry angle. This angle can expand or compress the diffracted beam size relative to the incoming size. This angle also can alter the diffraction properties of the crystal and has an impact on the Laue geometry and discussion of this angle is a significant part of this manuscript.

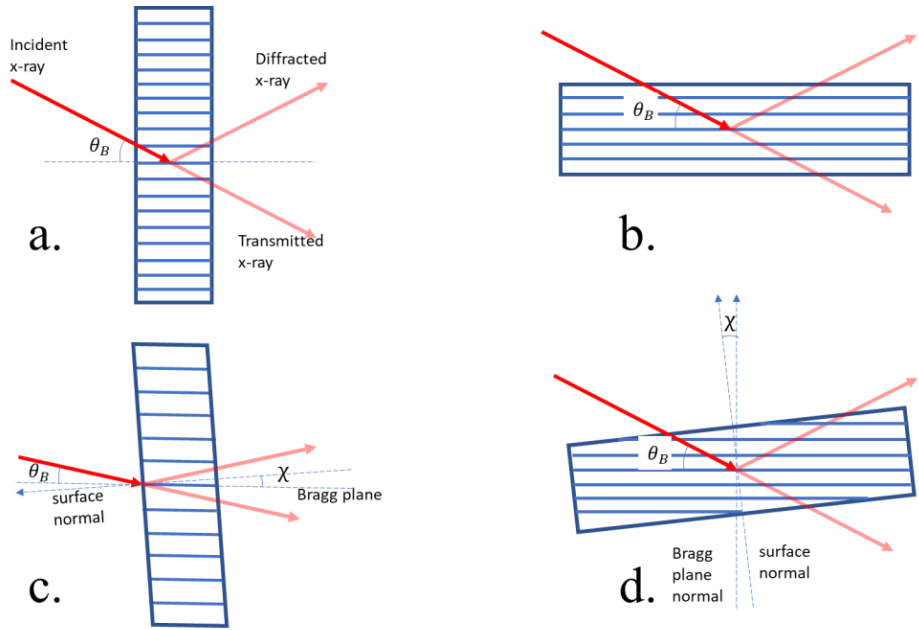

Figure 1. Crystal plate diffraction geometries. Figures. a & c show the transmission or Laue geometry and figures. b & d show the reflection or Bragg geometry. The figures a & b are the symmetric cases and the figures c & d are the asymmetric cases where the lattice planes are inclined by an asymmetry angle $\chi$ relative to the crystal plate surface.

The properties of single crystals in both geometries are well understood [1–3]. Often, there is a need to focus or increase the energy dispersion so elastic bending of crystals is often desired. The diffraction properties of elastically bent Laue crystals can be described by wave theories [4–6] and lamella models [7]. For the discussion that follows, the crystals will be elastically bent single crystals such as silicon.

Suppose, for simplicity, that a focusing bent Laue crystal is considered with a polychromatic incident x-ray beam. Similar arguments will apply for a defocusing or virtual focus system. The focal properties of such a crystal is complicated by two effects. One, there is a geometric focus arising from the overall curvature of the lattice in the diffraction plane. Also, each ray that intercepts the crystal will generate a bundle of diffracted rays that will focus (either a real or virtual focus). An example of the geometric focusing arrangement is shown in Fig. 2a and the effect of diffraction on each ray in Fig. 2b.

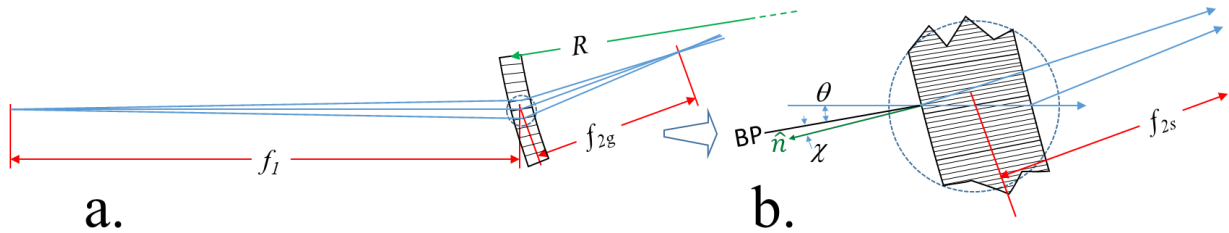

Figure 2. Bent asymmetric Laue focusing. Figure a shows the geometric focusing with the source at a distance $f_1$, the bent Laue crystal with radius R, asymmetry angle $\chi$, Bragg angle $\theta$ and focal distance $f_{2g}$. Figure b shows a chosen relationship between the incident ray in relation to the Bragg planes (BP), the crystal surface normal ($\hat{n}$) and the asymmetry angle $\chi$. The geometry shown results in rays diffracted at the entrance and exit of the crystal converging and focusing at $f_{2s}$.

Normally, the two foci do not coincide with each other. A carefully chosen asymmetry angle is critical for matching the two. It has been previously reported [8] that coinciding the geometric focus and single-ray focus optimizes the focal size. Lately, it has been also reported that when the coincidence condition is met, the energy resolution of the focused beam is optimized as well [3,9]. This condition has been referred to as the 'magic condition' [9], and it has enabled several novel imaging techniques, e.g. spectral K-edge subtraction imaging [9,10], wide field imaging - energy dispersive x-ray absorption speciation, and phase preserving beam expander [3]. Principle terms needed for designing a bent Laue crystal that match the magic condition are asymmetry angle $\chi$, Bragg angle for the center energy of interest, $\theta_B$, crystal bending radius, $R$, Poisson's ratio, $\nu$, and source to crystal distance $f_1$.

The presented work explains conceptually and mathematically how the 'magic condition' is met for a bent Laue crystal.

## 2. CONCEPTUAL VIEW OF THE MAGIC CONDITION

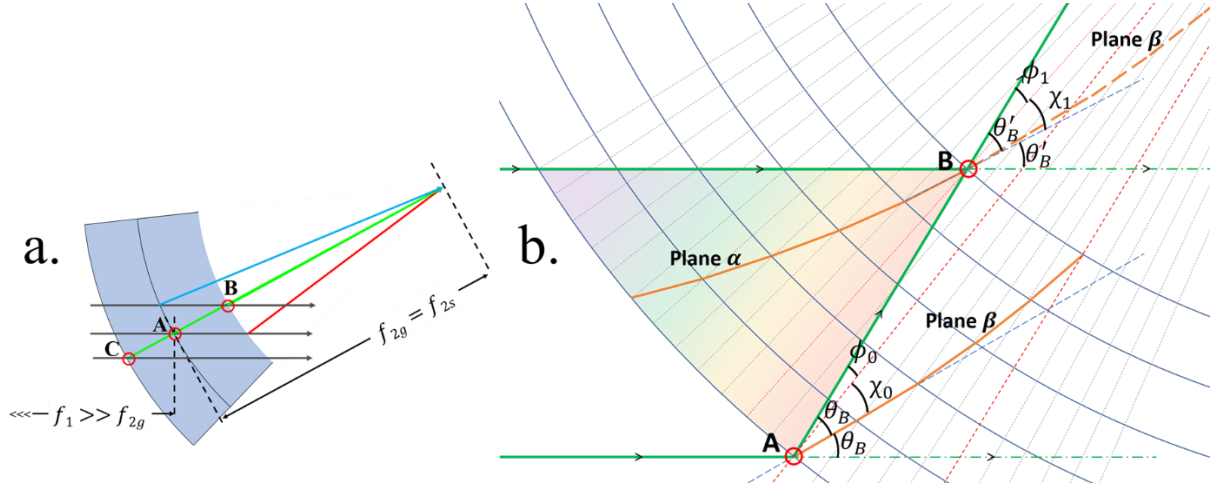

Figure 3. Energy dispersion of a bent Laue crystal at the 'magic condition'. Figure a shows the condition where the geometric and single-ray focus match. Figure b shows the detailed view of the relationship between bending of lattice planes α and β, local Bragg and asymmetry angles at the two locations A and B.

The magic condition ($f_{2g} = f_{2s}$) is illustrated in Fig. 3a, where the x-ray beams diffracted at point $A$, $B$ and C go in the same direction. Point $A$ is in the neutral plane of the crystal which is defined as the reference plane that separates the unbent crystal thickness into equal halves. The central plane is neither compressed nor expanded. Points B and C are the exit and entrance points on the crystal surface located on the concave and convex side of the crystal, respectively. As shown in Fig. 3a, the concave side of the crystal is compressed while the convex side is expanded.

Figure 3b gives a detailed view of the compressed half of the crystal. The magic condition $f_{2g} = f_{2s}$ implies that the diffraction angle at points $A$ and $B$ are the same, or $\theta_B = \theta'_B$. According to Bragg's law, when $\theta_B = \theta'_B$, the energy of the x-ray diffracted at the right end of lattice plane $\alpha$ is the same as the energy of the one diffracted at the left end of lattice plane $\beta$. Since the effective Bragg angle on lattice plane $\alpha$ and lattice plane $\beta$ is increasing along the horizontal direction and $\theta_B$ equals $\theta'_B$, lattice plane $\beta$ then acts like an extension of lattice plane $\alpha$ at point $B$. And from equation 9 (in section 3.4), we know that when this dispersive condition is met, the thickness of the crystal, $T$, is not part of the equation. This means at every point in the crystal along the way of the diffracted x-ray, the lattice has the same diffraction angle as the exit point $B$. Every diffracted x-ray on the left side of this path has lower energy, and every diffracted x-ray on the right side of this path has higher energy as indicated by the colors in Fig. 3a. So that lattice planes $\alpha$ and $\beta$ are appearing as a single lattice plane where point $A$ and $B$ are stitched together. When this condition is met, not only the lattice planes $\alpha$ and $\beta$, but all lattice planes in the bent Laue crystal, act like bigger Bragg reflection planes overlapping everyone else with their Bragg angles perfectly aligned.

## 3. MATHEMATICAL VIEW OF THE MAGIC CONDITION

To simplify the discussion, we take the concave compressed half of the crystal as an example. From the crystal geometry shown in Fig. 3b, one can easily tell $\theta_B$ equals the sum of $\chi$ and $\phi$, where $\chi$ is the angle between the diffraction plane and the surface normal (the asymmetry angle) and $\phi$ is the angle between the x-ray diffraction direction and the crystal surface normal. The condition that $\theta_B = \theta'_B$ can be described as the sum of the changes of $\chi$ and $\phi$ equaling zero.

$$\Delta\chi + \Delta\phi = (\chi_1 - \chi_0) + (\phi_1 - \phi_0) = 0 \tag{1}$$

The change of $\chi$ angle is caused by the deformation of the crystal which is, for example, on the concave side of the crystal, the compression of the crystal in one dimension and the resulting elastic expansion in the two transverse dimensions. The change of $\phi$ angle can be determined with the magic condition geometry. Details will be discussed in section 3.3.

## 3.1 Thickness of a deformed crystal

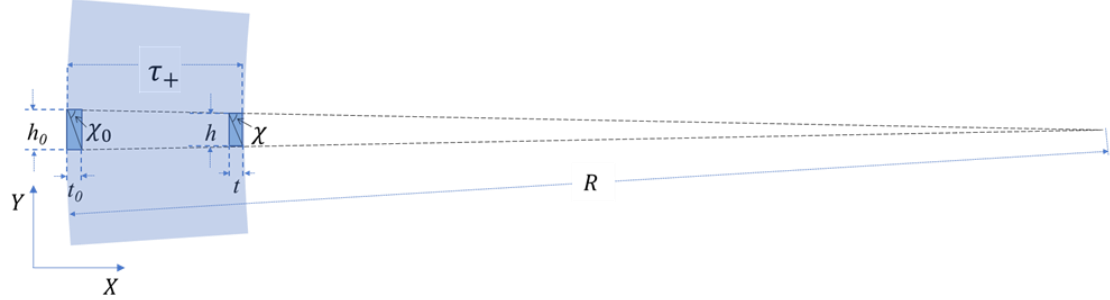

Figure 4. Schematic to estimate the change in bent crystal thickness and asymmetric angle $\chi$. The relevant dimensions and angles are discussed in the text.

Due to the bending force, the crystal is compressed on the concave side and expanded on the convex side in the direction of axis $Y$ as shown in Fig. 4. The compression or expansion causes an opposite deformation in the direction of axis $X$ and $Z$. Axis $Z$ (not shown in the figure) is orthogonal to the plane of $XY$, and not considered in the discussion here.

The thickness of the crystal (along axis $X$) is changed by bending considering the elastic property of the crystal through Poisson's ratio, which is defined by

$$\nu = -(\Delta X/X)/(\Delta Y/Y)$$

The negative sign implies that the compression (expansion) $\Delta Y$ results in a transverse expansion (compression) $\Delta X$. Thus, a crystal when bent will experience a growth in thickness on the compressed concave side and a reduction in thickness on the expanded convex side.

It can be shown that the thickness $\tau$ of bent crystal can be related to the unbent distance, $x$, away from the neutral plane along $X$ direction, by

$$\tau = \left(e^{\frac{\nu x}{R}} - 1\right) \cdot \frac{R}{\nu} \tag{2}$$

where $R$ is the bending radius of the crystal plate. As a value, $\tau$ and $x$ will be positive on the compressed side and negative on the expanded side. $\tau$ will be used in the discussion on $\Delta\phi$ and $\Delta\chi$ because using this variable avoids accounting for the change of sign when crossing the neutral plane.

For bent crystal with original thickness $T$, the thickness values on each side of the neutral planes are

$$\tau_\pm = \pm\left(e^{\pm\frac{\nu T/2}{R}} - 1\right) \cdot \frac{R}{\nu} \tag{3}$$

where the plus sign refers to the compressed side of the crystal and the minus sign refers to the expanded side. Note that $\tau_+$ and $\tau_-$ are both positive values. Therefore, $\tau = \tau_+$ and $\tau = -\tau_-$ for the compressed and expanded sides, respectively.

Since the original thicknesses for the compressed half of the crystal and the expanded half of the crystal are both $T/2$, the total thickness $T_{bent}$ after bending with radius $R$ is, using equation 3,

$$T_{bent} = \tau_+ + \tau_- = \frac{2R}{\nu} \cdot \sinh\frac{\nu T}{2R} \tag{4}$$

In the limit of large bending radius ($\nu T/2R \to 0$), the bent crystal thickness is just flat crystal thickness $T$, as expected.

## 3.2 The change in the angle between the diffraction plane and the surface normal, $\Delta\chi$

The change in $\chi$ over the crystal thickness is also a result of the deformation of crystal as shown in Fig. 4.

Now that we have $T_{bent}$, we can calculate $h$ and $t$ for determining $\tan(\chi + \Delta\chi)$. Assuming in the first unit cell (the left dark blue shaded box in Fig. 4), $\tan\chi_0 = \frac{h_0}{t_0}$, then in the last unit cell (the right blue box),

$$h = h_0 \cdot (1 - \tau_+/R)$$

$$t = t_0 \cdot (1 + \nu \cdot \tau_+/R) \,.$$

Therefore,

$$\tan(\chi + \Delta\chi) = \frac{h}{t} = \frac{1 - \tau_+/R}{1 + \nu \cdot \tau_+/R} \cdot \frac{h_0}{t_0} = \frac{1 - \tau_+/R}{1 + \nu \cdot \tau_+/R} \cdot \tan\chi_0.$$

Replacing $\tan(\chi_0 + \Delta\chi)$ with $\frac{\tan\chi_0 + \tan\Delta\chi}{1 - \tan\chi_0 \cdot \tan\Delta\chi}$, then

$$\frac{\tan\chi_0 + \tan\Delta\chi}{1 - \tan\chi_0 \cdot \tan\Delta\chi} = \frac{1 - \tau_+/R}{1 + \nu \cdot \tau_+/R} \cdot \tan\chi_0.$$

Solving the equation using the small angle approximation for $\Delta\chi$, we get

$$\Delta\chi_+ = \tan\Delta\chi = \frac{-(1 + \nu) \cdot \tau_+ \cdot \tan\chi_0}{(R + \nu\tau_+) + (R - \tau_+) \cdot \tan^2\chi_0}$$

It can be generalized to both side of the neutral plane as

$$\Delta\chi = \tan\Delta\chi = \frac{-(1+\nu)\cdot\tau\cdot\tan\chi_0}{(R+\nu\tau)+(R-\tau)\cdot\tan^2\chi_0} \tag{5}$$

The term $\tau$ can be replaced by the equation 2 for the calculation with flat crystal thickness if needed.

The total $\Delta\chi_T$ across the crystal is then

$$\Delta\chi_T = \frac{-(1+\nu)\cdot\tau_+\cdot\tan\chi_0}{(R+\nu\tau_+)+(R-\tau_+)\cdot\tan^2\chi_0} - \frac{(1+\nu)\cdot\tau_-\cdot\tan\chi_0}{(R-\nu\tau_-)+(R+\tau_-)\cdot\tan^2\chi_0}$$

$$\approx -\frac{\tau_+ + \tau_-}{2R}(1+\nu)\sin 2\chi_0$$

$$= -\frac{T_{bent}}{2R}(1+\nu)\sin 2\chi_0 \quad \text{or} \quad -\frac{T_{bent}}{R}(1+\nu)\frac{\tan\chi_0}{1+\tan^2\chi_0} \tag{6}$$

## 3.3 The change in the angle between the x-ray diffraction direction and the crystal surface normal, $\Delta\phi$

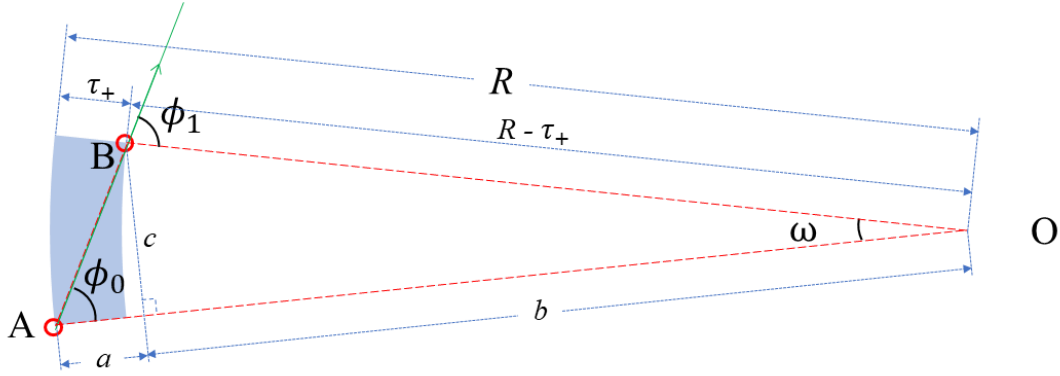

Figure 5. Schematic to estimate the change in $\phi$. Point A and B are the same as in Fig. 3. The relevant dimensions and angles are discussed in the text.

Figure 5 shows again the compressed half of the crystal and $O$ is the center of the crystal bending circle. $AO$ and $BO$ are along the corresponding crystal surface normal directions, and $AB$ is the path of the diffracted single x-ray in the crystal. It is obvious that the change of angle $\phi$ is equal to angle $\omega$,

$$\omega = \phi_1 - \phi_0.$$

In the triangle $ABO$,

$$a + b = R,$$

where

$$a = c/\tan\phi_0 = c/\tan(\theta_B - \chi_0) = \frac{(R - \tau_+) \cdot \sin\omega}{\tan(\theta_B - \chi_0)},$$

$$b = (R - \tau_+) \cdot \cos\omega.$$

Therefore,

$$\frac{(R - \tau_+) \cdot \sin\omega}{\tan(\theta_B - \chi_0)} + (R - \tau_+) \cdot \cos\omega = R.$$

Using the small angle approximation, we have

$$\frac{(R - \tau_+) \cdot \omega}{\tan(\theta_B - \chi_0)} + (R - \tau_+) \cdot \left(1 - \frac{\omega^2}{2}\right) = R.$$

Solving the equation for $\omega$, and we get 2 solutions,

$$\omega_1 = \frac{1}{\tan(\theta_B - \chi_0)} - \sqrt{\frac{1}{[\tan(\theta_B - \chi_0)]^2} - \frac{2\tau_+}{R - \tau_+}}$$

and

$$\omega_2 = \frac{1}{\tan(\theta_B - \chi_0)} + \sqrt{\frac{1}{[\tan(\theta_B - \chi_0)]^2} - \frac{2\tau_+}{R - \tau_+}}$$

The first solution, $\omega_1$, is the solution we need, while $\omega_2$ exists mathematically but it is not practical for the optical application here.

Generalizing it to both sides of the neutral plane,

$$\Delta\phi = \frac{1}{\tan(\theta_B - \chi_0)} - \sqrt{\frac{1}{[\tan(\theta_B - \chi_0)]^2} - \frac{2\tau}{R-\tau}} \tag{7}$$

The total angler change $\Delta\phi_T$ across the entire crystal thickness is then

$$\Delta\phi_T = \sqrt{\frac{1}{[\tan(\theta_B - \chi_0)]^2} + \frac{2\tau_-}{R+\tau_-}} - \sqrt{\frac{1}{[\tan(\theta_B - \chi_0)]^2} - \frac{2\tau_+}{R-\tau_+}}$$

$$\approx \frac{\tau_+ + \tau_-}{R} \tan(\theta_B - \chi_0)$$

$$= \frac{T_{bent}}{R} \tan(\theta_B - \chi_0) \tag{8}$$

### 3.4 Calculation for the condition: $\Delta\chi_T + \Delta\phi_T = 0$

Adding the two angular contributions (equations 6 and 8) and forcing the sum to be zero results in,

$$\frac{T_{bent}}{R} \tan(\theta_B - \chi_0) - \frac{T_{bent}}{R}(1+\nu)\frac{\tan\chi_0}{1+\tan^2\chi_0} = 0$$

or

$$(2+\nu) \cdot \tan\chi_0 + \nu \cdot \tan\theta_B \cdot \tan^2\chi_0 + \tan^3\chi_0 - \tan\theta_B = 0 \tag{9}$$

Solving this equation for $\chi$ is best done numerically. There is an analytical solution, but it is not insightful. When applied to high energy x-rays, both $\theta_B$ and $\chi_0$ are close to zero. Using the approximations of $\tan^2\chi_0 \approx 0$ & $\tan^3\chi_0 \approx 0$, we then get

$$\tan\chi_0 = \frac{\tan\theta_B}{\nu+2} \tag{10}$$

which agrees with the approximated solution for magic condition from other literature[3].

## 4. ENERGY SPREAD OF THE SINGLE EXIT RAY

With a source at an infinite distance and assuming the lattice d-spacing is not affected by the crystal deformation, the magic condition would be the ultimate energy dispersion condition at a specific energy for a bent Laue crystal. However, the energy spread mostly will not be smaller than that allowed by an unbent crystal.

When the d-spacing variation caused by the crystal deformation is taken into account, the energy spread in the single exit ray is increased according to Bragg's law.

When $\chi_0 = 0$, the d-spacing on the compression side of the crystal $d' = d_0\left(1 - \frac{\tau}{R}\right)$.

When $\chi_0 \neq 0$,

$$d' = d_0\left(1 - \frac{\tau}{R}\right)\frac{\cos\chi'}{\cos\chi_0} \tag{11}$$

Although $d'$ is at the compressed side of the crystal, $d'$ is not necessarily smaller that $d_0$. When

$$\frac{R - \tau}{R} > \frac{\cos\chi_0}{\cos\chi'},$$

$d'$ is greater than $d_0$, which means the distance of the lattice planes are expanded rather than compressed at the compression side of the bent crystal. This condition can also be easily described with crystal original thickness and asymmetry angle.

The bandwidth $\Delta E/E$ from a bent crystal arising from d-spacing variation will be the difference between the d-spacing on the compressed side from the uncompressed side.

$$\Delta E/E = \Delta d/d = |(d_+ - d_-)/d_0|$$

The approximated first order value of the bandwidth is

$$\frac{\Delta d}{d_0} = \frac{T}{R}(\cos^2\chi_0 - \nu \sin^2\chi_0) \tag{12}$$

There are other aspects that will be involved in the energy resolution of such a monochromator. There is only one energy that matches the magic condition. The divergence of the beam in the diffraction plane will create energies that "wander off" the matching condition. Other factors will also affect the ultimate energy dispersive properties, for example, i) the spatial resolution of the detector, ii) the x-ray source size, iii) the intrinsic Darwin width of the effective Bragg plane, iv) the monochromatic beam spread on a detector[12]. When there is an asymmetry angle and the magic condition is met, it can be considered as the mono beam spread (the 4th term) is minimized. Every single-ray in the diffracted beam is monochromatic and the only adverse contribution of the crystal to the undesired energy spread in a monochromatic single-ray is caused by the d-spacing variation over the bent crystal.

## 5. CONCLUSIONS

A conceptual and mathematical model of magic condition bent Laue focusing geometry was presented with emphasis on its energy dispersive properties. The condition required for focusing and the energy dispersion are derived along with their approximate solutions. The solutions for the local $\chi$ and d-spacing values can be used to ray trace diffracted beams. This will allow us to better develop single or multiple crystal bent Laue optics for imaging and energy dispersive spectroscopy applications.

## ACKNOWLEDGEMENTS

Research was funded by the Natural Sciences and Engineering Research Council of Canada (PQ, DC) and support from the Canada Research Chairs (DC).